\documentstyle[11pt,amsfonts,axodraw]{article}
\newskip\humongous \humongous=0pt plus 1000pt minus 100pt

\newif\ifdtup

\newcounter{eqnumber}[section]
\renewcommand{\theeqnumber}{\thesection.\arabic{eqnumber}}
\def\equn{
\refstepcounter{eqnumber}
\eqno({\rm \theeqnumber})
}

\makeatletter
\def\@eqnnum{\hbox{\reset@font\rm(\theequation)}}
\let\make@eqnnum=\@eqnnum %
\def\eqnum#1{\dec@eqnnum \global\def\make@eqnnum{\reset@font\rm(#1)}%
\def\@currentlabel{#1}%
}
\def\inc@eqnnum{\addtocounter{equation}{1}}
\def\dec@eqnnum{\addtocounter{equation}{-1}}
\@definecounter{equation}%
\@addtoreset{equation}{section} %
\def\theequation@prefix{{\thesection}.} %
\def\theequation{\theequation@prefix\arabic{equation}}%
\makeatother

\def\eqn#1{eq.~(\ref{#1})}
\def\Eqn#1{Eq.~(\ref{#1})}
\def\eqns#1#2{eqs.~(\ref{#1}) and~(\ref{#2})}

\def\fig#1{fig.~{\ref{#1}}}
\def\Fig#1{Fig.~{\ref{#1}}}

\def\sec#1{section~{\ref{#1}}}

\def\app#1{appendix~\ref{#1}}

\catcode`@=11  

\def\qq{\qquad}
\def\qb{{\bar q}}
\def\tr{\mathop{\rm tr}\nolimits}
\def\si{\sigma}

\def\Gr{\mathop{\rm Gr}\nolimits}
\def\g{\gamma}

\def\Ord{{\cal O}}
\def\A{{\cal A}}

\def\OP{{\rm OP}}
\def\COP{{\rm COP}}

\def\hf{{\textstyle {1\over2}}}

\def\tree{{\rm tree}}
\def\nf{n_{\! f}}
\def\oneloop{{1 \mbox{-} \rm loop}}
\def\twoloop{{2 \mbox{-} \rm loop}}

\def\Re{{\rm Re}}

\begin{document}
\begin{titlepage}
\begin{flushright}

hep-ph/9910563 \hfill SLAC--PUB--8294\\
DFTT 53/99\\
October, 1999\\
\end{flushright}

\vskip 2.cm

\begin{center}
{\Large\bf New Color Decompositions for Gauge Amplitudes\\ 
at Tree and Loop Level}\\

\vspace{1.cm}

{Vittorio Del Duca$^1$}\\
\vspace{.2cm}
{\sl Istituto Nazionale di Fisica Nucleare\\ Sezione di Torino\\
via P. Giuria, 1\\ 10125 - Torino, Italy}\\

\vspace{.5cm}

{Lance Dixon$^2$}\\
\vspace{.2cm}
{\sl Stanford Linear Accelerator Center\\ 
Stanford University\\ 
Stanford, CA 94309, USA}\\

\vspace{.5cm}

{Fabio Maltoni}\\
\vspace{.2cm}
{\sl Dipartimento di Fisica Teorica\\
Universit\`a di Torino\\
via P. Giuria, 1\\ 10125 - Torino, Italy}\\
\vspace{.5cm}
\end{center}

\begin{abstract}
Recently, a color decomposition using structure constants was introduced
for purely gluonic tree amplitudes, in a compact form involving only the
linearly independent subamplitudes.  We give two proofs that this
decomposition holds for an arbitrary number of gluons.  We also present
and prove similar decompositions at one loop, both for pure gluon
amplitudes and for amplitudes with an external quark-antiquark pair.
\end{abstract}

\vskip 1cm
\begin{center}
{\sl Submitted to Nuclear Physics B}
\end{center}

\vfill
\noindent\hrule width 3.6in\hfil\break
${}^{1}$Address after November 1, 1999: Theory Division, CERN, CH 1211
Geneve 23, Switzerland\hfil\break
${}^{2}$Research supported by the US Department of Energy under grant 
DE-AC03-76SF00515\hfil\break
\end{titlepage}

\baselineskip 16pt


\section{Introduction}
\label{IntroSection}

The computation of multi-parton scattering amplitudes in QCD is essential
for quantitative predictions of multi-jet cross-sections at high-energy
colliders.  The Feynman diagrams for such an amplitude can generate a
thicket of complicated color algebra, tangled together with expressions
composed of kinematic invariants.  An extremely useful way to disentangle
the color and kinematic factors is via {\it color
decompositions}~\cite{ColorCLS,ColorIdentities,Color,MPReview,BKLoopColor}
in terms of Chan-Paton factors~\cite{ChanPaton}, or traces of $SU(N_c)$
matrices in the fundamental representation, $\lambda^a$.  The traces are
multiplied by purely kinematical coefficients, called partial amplitudes
or subamplitudes.  Originally motivated by the representation of gluon
amplitudes in open string theory, such decompositions have been widely
studied and applied at both the tree level~\cite{MPReview} and the loop
level~\cite{BDKReview}.  Two major advantages of the trace-based color
decompositions are that (a) subamplitudes are gauge-invariant, and (b) an
important `color-ordered' class of them receive contributions only from
diagrams with a particular cyclic ordering of the external partons; these
subamplitudes therefore have much simpler kinematic properties than the
full amplitude.

Although the standard color decompositions are very effective, in some
cases they are not quite optimal.  For example, the decomposition of the
purely gluonic tree amplitude is overcomplete.  Explicitly, the $n$-gluon 
amplitude is written in terms of $(n-1)!$ single-trace color structures, as
$$
\A_n^\tree(g_1,g_2,\ldots,g_n) = g^{n-2} \sum_{\si \in S_{n-1}} 
   \tr( \lambda^{a_1} \lambda^{a_{\si_2}} \cdots \lambda^{a_{\si_n}} )
   A_n^\tree(1,\si_2,\ldots,\si_n),
\equn\label{TreeGluonColorOld}
$$
where $A_n^\tree$ are the subamplitudes.  The cyclic invariance of the
trace has been used to fix its first entry, and the sum is over the set 
of non-cyclic permutations of $n$ elements, 
$\si \in S_{n-1} \equiv S_n/{\Bbb Z}_n$, where 
$\si(2,\ldots,n)=(\si_2,\ldots,\si_n)$.
However, the $(n-1)!$ subamplitudes appearing in the equation are not all
independent.  Besides cyclic and reflection invariances, which they
inherit from the traces, subamplitudes also obey a $U(1)$
decoupling identity (also called a dual Ward identity, or subcyclic 
identity)~\cite{ColorIdentities,BGRecRelations},
$$
A_n^\tree(1,2,3,\ldots,n) + A_n^\tree(1,3,2,\ldots,n) + \cdots + 
A_n^\tree(1,3,\ldots,n,2) = 0.
\equn\label{DecouplingId}
$$
This identity can be derived by setting $\lambda^{a_2}$ equal to the unit
matrix and collecting terms containing the same trace; this gives the
amplitude for a $U(1)$ `photon' and $n-1$ $SU(N_c)$ gluons, 
which must vanish.  More general identities can be derived by assigning
the $SU(N_c)$ generators for the external gluons to commuting subgroups 
such as $SU(N_1) \times SU(N_2)$~\cite{BGNewDecoupling}.  Kleiss and
Kuijf found a linear relation between subamplitudes~\cite{KK} 
(see \sec{TreeSection}), which is consistent with all of these identities,
and which reduces the number of linearly independent subamplitudes from 
$(n-1)!$ to $(n-2)!$.

In contrast, the conventional color factors for gluonic Feynman diagrams
are composed of structure constants $f^{abc}$.  In this
case, the $U(1)$ and generalized decoupling identities are all manifest; 
for example, the $U(1)$ generator commutes with all other generators, so
all structure constants containing it are zero.  
On the other hand, a particular string of contracted structure constants
will not appear {\it ab initio} with a gauge-invariant kinematic
coefficient.  To pass from the $f^{abc}$-based decomposition to the 
trace-based decomposition, one substitutes 
$f^{abc} = -i \tr \left( \lambda^a \lambda^b \lambda^c 
- \lambda^b \lambda^a \lambda^c \right)$ 
and simplifies and collects the various traces.  In the
process, gauge invariance of the partial amplitudes is gained, but
properties such as $U(1)$ decoupling are no longer manifest.  
Another disadvantage of trace-based decompositions at tree level is
that the color structure of amplitudes in the multi-Regge kinematics
is obscured~\cite{ptlip}.  

The trace-based decomposition is also not optimal at the loop level.  At
one loop, the standard color decomposition~\cite{BKLoopColor} for
$n$-gluon amplitudes includes double trace structures, $\tr(\lambda^{a_1}
\cdots \lambda^{a_{c-1}})$ $\times \, \tr(\lambda^{a_c} \cdots
\lambda^{a_{n}})$, in addition to single traces of the type that appear at
tree level.  The subamplitudes multiplying the single trace structures
give the leading contributions in the large $N_c$ limit, and they are
color-ordered.  The double-trace subamplitudes, corresponding to
subleading-in-$N_c$ contributions, can be written in terms of permutations
of the leading (color-ordered) subamplitudes~\cite{BDDKNeq4} in a formula
reminiscent of the tree-level Kleiss-Kuijf relation (see
\sec{OneLoopSection}).  Because the permutation formula is rather
complicated, its implementation in a numerical program can be rather slow.
Similar formulae hold for one-loop amplitudes with two external quarks and
$n-2$ gluons~\cite{TwoqNgluon}.

Recently, a new color decomposition for the $n$-gluon tree amplitude,
in terms of structure constants rather than traces, has been 
presented~\cite{dfm}, whose kinematic coefficients are just 
color-ordered subamplitudes.  Because of this property, the new 
decomposition retains all the advantages of trace-based decompositions, 
yet avoids the disadvantages mentioned above.  In particular, the 
tree-level $n$-gluon decomposition is automatically given in terms of the 
$(n-2)!$ independent subamplitudes, in a form that is very convenient 
for analyzing the high-energy limit.  

The purpose of this paper is to derive the new tree-level decomposition in
two different ways, and to present similar $f^{abc}$-based color
decompositions at one loop, whose kinematic coefficients are again
color-ordered subamplitudes.  The leading and subleading-in-$N_c$
contributions combine neatly into one expression in the new
decompositions. Also, gluons circulating in the 
loop are put on the same footing as fermions in the loop.

Parton-level cross-sections require amplitude squares or interferences
which are summed over all external color labels.  We provide general
expressions for color-summed cross-sections in terms of the new 
color decompositions.  In \app{sec:app1} we give explicit evaluations
for quantities encountered in cross-section computations through
$\Ord(\alpha_s^4)$.

The paper is organized as follows.  In \sec{TreeSection} we give two proofs
of the new color decomposition for the $n$-gluon tree amplitude
conjectured in ref.~\cite{dfm}, and provide accordingly the square
of the tree amplitude summed over colors. In \sec{OneLoopSection}
we present (and prove) new color decompositions for 
one-loop $n$-gluon amplitudes and one-loop amplitudes with an
external quark-antiquark pair plus $n-2$ gluons. In addition, we
compute the color-summed interference terms between tree amplitudes and
one-loop amplitudes, and the square of one-loop amplitudes,
which are relevant respectively for next-to-leading order (NLO) and
next-to-next-to-leading order (NNLO) calculations of jet production rates.
The color-summed interference terms and squares are given in terms of
color matrices; appendix A provides many of those
required for jet rate computations up to $\Ord(\alpha_s^4)$.
Finally, we outline how a new color decomposition for one-loop 
amplitudes with external photons, gluons and a quark-antiquark pair
can be obtained from the amplitudes with only gluons and a $q\qb$ pair.
In \sec{ConclusionsSection} we summarize our present understanding of 
the color decomposition of tree-level and one-loop amplitudes,
and comment briefly on possible extensions to multi-loop amplitudes.


\section{Tree Color Decomposition}
\label{TreeSection}

The new color decomposition for the $n$-gluon tree amplitude 
is\footnote{We choose the normalization of the fundamental representation 
matrices as $\tr(\lambda^a \lambda^b) = \delta^{ab}$.  Hence our structure
constants $f^{abc}$ are larger than the conventional ones by a factor 
of $\sqrt{2}$.}~\cite{dfm}
\begin{eqnarray}
 \A_n^\tree (g_1,\ldots,g_n)&=&    
 (ig)^{n-2} \sum_{\si\in S_{n-2}}
      f^{a_1 a_{\si_2} x_1} f^{x_1 a_{\si_3} x_2} \cdots f^{x_{n-3} 
a_{\si_{n-1}} a_n}
    A_n^\tree(1,\si_2,\ldots,\si_{n-1},n)\, ,\nonumber\\
&=& g^{n-2} \sum_{\si\in S_{n-2}}
      (F^{a_{\si_2}} \cdots F^{a_{\si_{n-1}}})_{a_1 a_n}
      A_n^\tree(1,\si_2,\ldots,\si_{n-1},n)\, ,
\label{GluonDecompNew}
\end{eqnarray}
where $(F^a)_{bc}\equiv i f^{bac}$ is an $SU(N_c)$ generator in the
adjoint representation.  This color decomposition is 
analogous to the standard decomposition for the tree amplitude with a
quark-antiquark pair and $n-2$ gluons~\cite{Color,MPReview},
\begin{eqnarray}
 \A_n^\tree (q_1,g_2,\ldots,g_{n-1},\qb_n)&=&    
 g^{n-2} \sum_{\si\in S_{n-2}}
  (\lambda^{a_{\si_2}} \cdots \lambda^{a_{\si_{n-1}}})^{~\bar{i}_n}_{i_1~}
   A_n^\tree (1_q,\si_2,\ldots,\si_{n-1},n_{\qb})\,.
\label{QuarkDecompOld}
\end{eqnarray}
The only difference between the two is the representation used
for the color matrices, namely the adjoint representation for the 
$n$-gluon amplitude and the fundamental representation for the amplitude
containing a ${\bar q} q$ pair.  

\Eqn{GluonDecompNew} is written in terms of the $(n-2)!$  subamplitudes 
where legs $1$ and $n$ are adjacent.  This is precisely the basis
of linearly independent subamplitudes which is singled out by the
Kleiss-Kuijf relation.  In fact, we shall show that \eqn{GluonDecompNew}
is equivalent to the Kleiss-Kuijf relation~\cite{KK}, which can be 
written as,
\begin{eqnarray}
A_n^\tree(1,\{\alpha\},n,\{\beta\}) =
 (-1)^{n_\beta} \sum_{\si\in \OP\{\alpha\}\{\beta^T\}}
 A_n^\tree(1,\si(\{\alpha\}\{\beta^T\}),n).
\label{KKequation}
\end{eqnarray}
Here $\{\alpha\} \cup \{\beta\} = \{2,3,\ldots,n-1\}$, $n_\beta$ is the
number of elements in the set $\{\beta\}$, the set $\{\beta^T\}$ is
$\{\beta\}$ with the ordering reversed, and $\OP\{\alpha\}\{\beta^T\}$ is
the set of `ordered permutations' (also called mergings~\cite{KK}) of the 
$n-2$ elements of $\{\alpha\} \cup \{\beta^T\}$ that preserve the 
ordering of the $\alpha_i$ within $\{\alpha\}$ and of the $\beta_i$ 
within $\{\beta^T\}$, while allowing for all possible relative orderings 
of the $\alpha_i$ with respect to the $\beta_i$.  

We wish to show that the new color decomposition~(\ref{GluonDecompNew}) 
is equivalent to inserting the Kleiss-Kuijf relation~(\ref{KKequation})
into the standard color decomposition~(\ref{TreeGluonColorOld}).
We first substitute for the structure constants appearing in 
\eqn{GluonDecompNew}, $f^{abc} = -i \tr(\lambda^a [\lambda^b,\lambda^c])$,
and use the identity
\begin{equation}
f^{a_1 a_2 x_1} f^{x_1 a_3 x_2} \cdots f^{x_{n-3} a_{n-1} a_n} =
(-i)^{n-2}\, {\rm tr}\left(\lambda^{a_1}\,\left[\lambda^{a_2},
\left[\lambda^{a_3},\ldots,\left[\lambda^{a_{n-1}}, \lambda^{a_n}
\right]\ldots\right]\right]\right).
\label{fpro}
\end{equation}
We want to identify all terms that contain a trace of the form
\begin{equation}
\tr(\lambda^{a_1} \lambda^{a_{\alpha_1}} \cdots \lambda^{a_{\alpha_{n-2-q}}} 
\lambda^{a_n} \lambda^{a_{\beta_1}} \cdots \lambda^{a_{\beta_q}}),
\label{DesiredTrace}
\end{equation}
because these should give rise to $A_n^\tree(1,\{\alpha\},n,\{\beta\})$.
Since there are $n-2$ commutators, there are $2^{n-2}$ terms on the 
right-hand side of \eqn{fpro}, but only $({n-2 \atop q})$ of them have 
exactly $q$ $\lambda$ matrices appearing to the right of $\lambda^{a_n}$.  
The $a_i$ indices on the matrices to the right of $\lambda^{a_n}$ must 
come in reversed order compared to how they appear in the $f^{abc}$ string,
but they can appear in that string in any relative order with respect
to the $a_i$ that end up to the left of $\lambda^{a_n}$.  Thus, for any 
ordered permutation $\si \in \OP\{\alpha\}\{\beta^T\}$, the subamplitude
$A_n^\tree(1,\si(\{\alpha\}\{\beta^T\}),n)$ appears in the new
decomposition~(\ref{GluonDecompNew}) accompanied by the desired
trace~(\ref{DesiredTrace}); a relative sign of 
$(-1)^q = (-1)^{n_\beta}$ comes from the commutators.
Collecting all such ordered permutations, we obtain the Kleiss-Kuijf
formula for $A_n^\tree(1,\{\alpha\},n,\{\beta\})$, \eqn{KKequation},
thus establishing its equivalence to the new color
decomposition~(\ref{GluonDecompNew}).

Hence we can derive the new $n$-gluon color decomposition via its
connection to the Kleiss-Kuijf relation.  The latter relation was checked
up to $n=8$ in ref.~\cite{KK}.  It can be proved for all $n$ using the
same techniques that were used to prove an analogous one-loop
formula~\cite{BDDKNeq4}, \eqn{sublanswer} below.  Consider first the set of
color-ordered Feynman diagrams with only three-point vertices, which are
`multi-peripheral' with respect to $1$ and $n$, by which we mean that
all other external legs connect directly to the line extending from $1$ to
$n$; i.e. there are no non-trivial trees branching off of this line.
Label the legs on one side of the $1$--$n$ line by $\alpha_i$, and those
on the other side by $\beta_j$.  These diagrams contribute to both sides
of the Kleiss-Kuijf relation~(\ref{KKequation}) in the proper way: There
is no relative ordering requirement on the $\alpha_i$ with respect to the
$\beta_j$ on the left-hand side of the relation, and the ordering is
summed over on the right-hand side; the sign factor $(-1)^{n_\beta}$ comes
from the antisymmetry of color-ordered three-point vertices (in Feynman
gauge).

Next consider diagrams with non-trivial trees attached to the $1$--$n$ line.
They work in exactly the same way as the multi-peripheral diagrams if the 
leaves (external legs) of each tree all belong to the same set, 
either $\{\alpha\}$ or $\{\beta\}$.  However, if a tree contains leaves 
from both sets, then the diagram does not appear on the left-hand side of
\eqn{KKequation}, so one must show that it cancels out of the permutation
sum on the right-hand side.  For trees with only three-point vertices,
this can be done using the antisymmetry of the vertices.  For other cases
the cancellations are slightly more complicated; one way of establishing
them is via the generalized decoupling identities~\cite{BGNewDecoupling}
mentioned in the introduction~\cite{BDDKNeq4}.

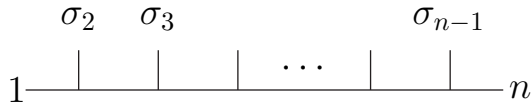
\begin{figure}[ht]
\begin{center}
\begin{picture}(200,60)(0,0)
\Text(10,15)[r]{{\Large $1$}}
\Line(10,15)(190,15) 
\Line(30,15)(30,30) \Text(30,40)[b]{{\Large $\si_2$}}
\Line(60,15)(60,30) \Text(60,40)[b]{{\Large $\si_3$}} 
\Line(90,15)(90,30) 
     \Text(115,20)[b]{{\Large $\cdots$}} 
\Line(140,15)(140,30)
\Line(170,15)(170,30) \Text(170,40)[b]{{\Large $\si_{n-1}$}} 
\Text(193,15)[l]{{\Large $n$}}
\end{picture}
\end{center}
\caption[a]{\small Graphical representation of a multi-peripheral color
factor.  A vertex stands for $f^{abc}$, and a bond for $\delta^{ab}$.}
\label{MultiPeriphFigure}
\end{figure}

\begin{figure}[ht]
\begin{center}
\begin{picture}(150,60)(0,0)
\Line(10,10)(20,20) \Line(10,50)(20,40)   \Line(20,20)(20,40)
\Line(20,20)(30,10) \Line(20,40)(30,50)
         \Text(40,30)[l]{{\Large $=$}}
\Line(60,20)(70,30) \Line(60,40)(70,30)   \Line(70,30)(90,30)
\Line(90,30)(100,20) \Line(90,30)(100,40)
         \Text(110,30)[l]{{\Large $-$}}
\Line(130,20)(140,30) \Line(130,40)(148,34)  \Line(152,32.6666)(160,30)   
\Line(140,30)(160,30)   \Line(160,30)(170,20) \Line(140,30)(170,40)
\end{picture}
\end{center}
\caption[a]{\small The Jacobi identity for structure constants in 
graphical notation.}
\label{JacobiFigure}
\end{figure}
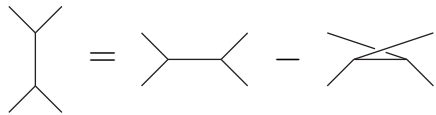

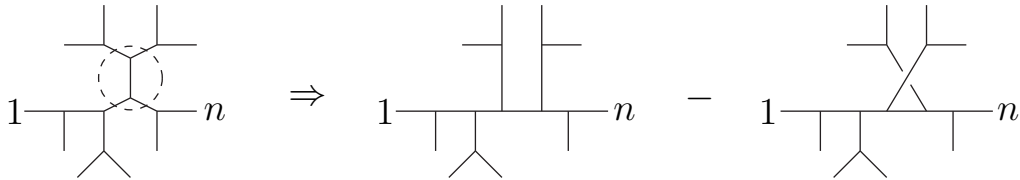
\begin{figure}[ht]
\begin{center}
\begin{picture}(400,95)(0,0)
\Text(10,40)[r]{{\Large $1$}}
\Line(10,40)(40,40) \Line(25,25)(25,40) \Line(40,25)(40,40)
\Line(40,25)(30,15) \Line(40,25)(50,15)
\Line(40,40)(50,45) \Line(50,45)(60,40) 
\Line(50,45)(50,60)
\Line(50,60)(40,65) \Line(50,60)(60,65) 
\Line(40,65)(25,65) \Line(40,65)(40,80)
\Line(60,65)(75,65) \Line(60,65)(60,80)
\Line(60,40)(60,25) \Line(60,40)(75,40)
\Text(78,40)[l]{{\Large $n$}}
\DashCArc(50,52.5)(12,0,360){3.1}
          \Text(110,45)[l]{{\Large $\Rightarrow$}}
\Text(150,40)[r]{{\Large $1$}}
\Line(150,40)(230,40)
\Line(165,25)(165,40) \Line(180,25)(180,40)
\Line(180,25)(170,15) \Line(180,25)(190,15)
\Line(190,40)(190,80) \Line(190,65)(175,65)
\Line(205,40)(205,80) \Line(205,65)(220,65)   \Line(215,40)(215,25)
\Text(233,40)[l]{{\Large $n$}}
          \Text(260,45)[l]{{\Large $-$}}
\Text(295,40)[r]{{\Large $1$}}
\Line(295,40)(375,40)
\Line(310,25)(310,40) \Line(325,25)(325,40)
\Line(325,25)(315,15) \Line(325,25)(335,15)
\Line(335,65)(335,80) \Line(335,65)(320,65)
\Line(335,40)(350,65)    \Line(335,65)(341,55) \Line(344,50)(350,40)
\Line(350,65)(350,80) \Line(350,65)(365,65)   \Line(360,40)(360,25)
\Text(378,40)[l]{{\Large $n$}}
\end{picture}
\end{center}
\caption[a]{\small Graphical representation of the color factor for a 
generic tree-level Feynman diagram.  Also shown is a step in its 
conversion to multi-peripheral form, by using the Jacobi identity in the
dashed region.}
\label{TreeSimpFigure}
\end{figure}

There is another way to prove \eqn{GluonDecompNew}.  For this argument it
is convenient to use a graphical notation for color factors made out of
structure constants, in which $f^{abc}$ is represented by a three-vertex
and $\delta^{ab}$ (an index contraction) is represented by a
line~\cite{ColorCLS}.  Then the color factors appearing in
\eqn{GluonDecompNew} are associated with the multi-peripheral color
diagrams shown in \fig{MultiPeriphFigure}.  The color factor for a generic
$n$-gluon Feynman diagram is not of this form.  However, we can use the
Jacobi identity
\begin{equation}
f^{dac} f^{cbe} - f^{dbc} f^{cae} = f^{abc} f^{dce},
\label{JacobiIdent}
\end{equation}
shown graphically in \fig{JacobiFigure}, to put it into this
form~\cite{ColorCLS}.  \Fig{TreeSimpFigure} represents the color factor
for a generic Feynman diagram graphically, as a tree structure with only
three-point vertices.  It also shows how it can be simplified into
multi-peripheral form by repeated application of the Jacobi identity.  (If
the line running from $1$ to $n$ is in the fundamental representation,
then the same steps, but using $\lambda^a \lambda^b - \lambda^b \lambda^a
= i f^{abc} \lambda^c$, lead directly to the color
decomposition~(\ref{QuarkDecompOld}) for a quark-antiquark pair and $n-2$
gluons.)

\begin{figure}[ht]
\begin{center}
\begin{picture}(450,60)(0,0)
\Line(3.432,20)(116.568,20) 
\Line(15,20)(15,49.8431) \Line(30,20)(30,55.981) \Line(45,20)(45,59.047)
\Line(60,20)(60,60) \Line(75,20)(75,59.047) \Line(90,20)(90,55.981)
\Line(105,20)(105,49.8431)
\SetWidth{1}
\Oval(60,30)(30,60)(0)
\ArrowArcn(60,30)(30,270.1,269.9)
\SetWidth{0.5}
   \Text(130,30)[l]{$=\ N_c^n + \Ord(N_c^{n-2})$}
\Line(253.432,20)(366.568,20) 
\Line(265,20)(280,55.981) \Line(280,20)(295,59.047) 
\Line(295,20)(286,28.9529)\Line(282,32.9320)(276,38.9006) 
          \Line(272,42.8797)(265,49.8431)
\Line(310,20)(310,60) \Line(325,20)(325,59.047) \Line(340,20)(340,55.981)
\Line(355,20)(355,49.8431)
\SetWidth{1}
\Oval(310,30)(30,60)(0)
\ArrowArcn(310,30)(30,270.1,269.9)
\SetWidth{0.5}
\Text(380,30)[l]{$=\ \Ord(N_c^{n-2})$}
\end{picture}
\end{center}
\caption[a]{\small Contraction of different multi-peripheral color factors
with a single $\lambda$ trace.  The heavy line with an arrow denotes
the fundamental representation.  The left diagram shows the 
contraction where $a_i$, $i=2,3,\ldots,n-1$, appear in the same order
in the $\lambda$ trace as in the multi-peripheral color factor.
Every other contraction gives rise to a nonplanar (and hence
color-suppressed) diagram of the type shown on the right.}
\label{NewColorOrthog}
\end{figure}
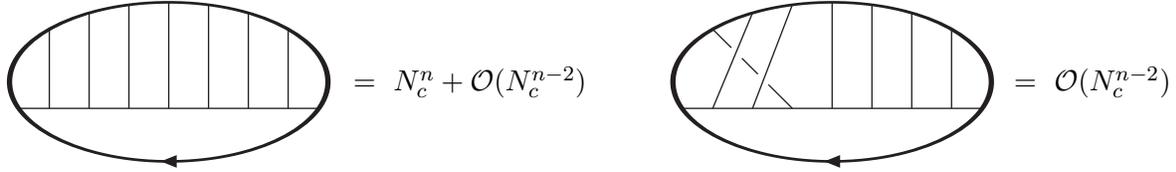

So far we have established that a color decomposition of the
form~(\ref{GluonDecompNew}) exists, but we have not yet shown that the
kinematic coefficients are equal to the partial amplitudes $A_n^\tree
(1,\si_2,\ldots,\si_{n-1},n)$.  However, we can do this rather simply by
equating the new decomposition (but with unknown kinematic coefficients)
to the standard one, \eqn{TreeGluonColorOld}, and contracting both sides
with $\tr(\lambda^{a_1} \lambda^{a_{\si_2}} \cdots \lambda^{a_{\si_{n-1}}}
\lambda^{a_n})$, i.e. summing over all $a_i$, $i=1,\ldots,n$, for some
permutation $\si \in S_{n-2}$.  Because the kinematic coefficients contain
no $N_c$ dependence, we may retain only the leading contraction terms as
$N_c \to \infty$.  But it is well known that single traces for different
cyclic orderings are orthogonal at large $N_c$~\cite{MPReview}.
Similarly, \fig{NewColorOrthog} shows that only one of the
multi-peripheral color factors in~\eqn{GluonDecompNew} survives the
contraction, because the other $S_{n-2}$ permutations give rise to
nonplanar, and hence $N_c$-suppressed, color topologies.
Thus the contraction selects a unique term from either side of the
equation, corresponding to the same ordering of the color indices $a_i$,
and with the same weight at large $N_c$, namely $N_c^n$.
This proves that the coefficients in the new color
decomposition~(\ref{GluonDecompNew}) are indeed the subamplitudes
$A_n^\tree(1,\si_2,\ldots,\si_{n-1},n)$.

Because we showed that the new color decomposition is equivalent to the 
Kleiss-Kuijf relation, our second proof of the new decomposition 
can alternatively be viewed as a proof of the Kleiss-Kuijf relation.
Interestingly, this proof does not rely on any kinematic properties of 
color-ordered Feynman diagrams; only group-theoretic properties such as 
the Jacobi identity and the large-$N_c$ behavior of color traces were used.

The decomposition~(\ref{GluonDecompNew}) has a color-ladder structure
which naturally arises in the configurations where the gluons are strongly
ordered in rapidity, i.e. in the multi-Regge kinematics~\cite{fkl}.
(Indeed, this is how \eqn{GluonDecompNew} was first
discovered~\cite{dfm}.)  In contrast, some laborious work is necessary to
obtain the color-ladder structure from the trace-based
decomposition~(\ref{TreeGluonColorOld}) in the multi-Regge
kinematics~\cite{ptlip}. Let gluons $2,\ldots,n-1$ be the particles produced
in a scattering where gluons 1 and $n$ are the incoming particles.
The available final-state phase space may be divided into $(n-2)!$
simplices, in each of which the produced gluons are ordered in rapidity.
Let us take the simplex defined by the rapidity ordering
$y_2 > y_3 > \cdots > y_{n-2} > y_{n-1}$, and consider its sub-simplex with
strong rapidity ordering $y_2 \gg y_3
\gg \cdots \gg y_{n-2} \gg y_{n-1}$.  Then the crucial point to note is that
the dominant subamplitudes in \eqn{TreeGluonColorOld} in the multi-Regge
limit are $A_n^\tree(1,\{\alpha\},n,\{\beta^T\})$, where $\{\alpha\}$ and
$\{\beta\}$ are both increasing sequences, whose union is
$\{2,3,\ldots,n-1\}$ ($\{\beta^T\}$ is $\{\beta\}$ in reversed order).  
In other words, $\OP\{\alpha\}\{\beta\}$ contains the sequence
$\{2,3,\ldots,n-1\}$.  If $q$ is the number of elements in $\{\beta\}$,
then for each $q$ there are $({n-2 \atop q})$ such choices of
$\{\alpha\}$, $\{\beta\}$.  Summing over $q=0,1,2,\ldots,n-2$, there are
$2^{n-2}$ orderings in total~\cite{lego}.  Using the identity~(\ref{fpro})
and the fact that in the multi-Regge kinematics
\begin{equation}
A_n^\tree(1,\{\alpha\},n,\{\beta^T\}) = (-1)^q A_n^\tree(1,2,\ldots,n),
\end{equation}
one obtains the color-ladder structure of \eqn{GluonDecompNew},
with just one allowed string of structure constants $f^{abc}$.
The procedure can be repeated for all of the $(n-2)!$ simplices,
thus generating the $(n-2)!$ strings of $f^{abc}$'s of \eqn{GluonDecompNew}.

The tree-level partonic scattering cross-section is given by the square of
the tree amplitude $\A_n^\tree$, summed over all colors.  This expression
can be written either of two ways,
\begin{eqnarray}
\sum_{\rm colors} |\A_n^\tree(1,\dots,n)|^2 
&=& (g^2)^{n-2} \sum_{i,j=1}^{(n-1)!} 
       c_{ij} A_i^\tree (A_j^\tree)^* \, \label{square1}\\
&=& (g^2)^{n-2} \sum_{i,j=1}^{(n-2)!} 
\tilde{c}_{ij} A_i^\tree (A_j^\tree)^* \, , \label{square2}
\end{eqnarray}
where \eqn{square1} is obtained from \eqn{TreeGluonColorOld}, while 
\eqn{square2} is obtained from \eqn{GluonDecompNew}, and the subscript
$i$ on $A_i^\tree$ now refers to the subamplitude $A_n^\tree$ evaluated 
for the  $i^{\rm th}$ permutation $P_i$ in $S_{n-1}$ or $S_{n-2}$, 
respectively. In \eqn{square1}, the color matrix $c_{ij}$ is 
\begin{eqnarray}
c_{ij} = 
\sum_{\rm colors} 
\tr(\lambda^{a_1} P_i \{ \lambda^{a_2} \cdots \lambda^{a_n} \} )
\left[ \tr(\lambda^{a_1} P_j \{ \lambda^{a_2} \cdots \lambda^{a_n} \} )
\right]^* ,
\label{cdef}
\end{eqnarray}
whereas in \eqn{square2} the matrix is
\begin{equation}
\tilde{c}_{ij} = \sum_{\rm colors} \left(P_i \{ F^{a_2} \cdots 
F^{a_{n-1}}\} \right)_{a_1 a_n}
\left[\left(P_j \{F^{a_2} \cdots F^{a_{n-1}}\}\right)_{a_1 a_n}
 \right]^*. \label{ctilde}
\end{equation}

In refs.~\cite{KK,BeGiKu90}, the Kleiss-Kuijf relation was used to
calculate $\tilde{c}_{ij}$ from $c_{ij}$.  From \eqn{ctilde} we see
that it can be calculated directly in terms of structure constants.
In \app{sec:app1} we give $\tilde{c}_{ij}$ for $n=4,5,6$. 

Although the expression~(\ref{square2}) seems to have fewer terms than
\eqn{square1}, the sparseness of the matrices is also important for 
determining which expression has the fastest numerical
evaluation~\cite{BeGiKu90}.  For an evaluation that is good only to 
leading order in $N_c$ --- the Leading Color Approximation, or LCA ---
it is best to use \eqn{square1} and the large-$N_c$ orthogonality of 
different single traces, to obtain
\begin{eqnarray}
\sum_{\rm colors} |\A_n^\tree(1,\dots,n)|^2 
&=& (g^2)^{n-2} {\cal C}_n(N_c)  \sum_{\si \in S_{n-1}} 
\left[ |A^\tree(1,\si_2,\ldots, \si_n)|^2 + \Ord\left(
\frac{1}{N_c^2}\right) \right] \, , \label{square1largeN}
\end{eqnarray}
where
\begin{eqnarray}
{\cal C}_n(N_c) = N_c^{n-2} (N_c^2-1) \, \label{calc}.
\end{eqnarray}
Up to $n=5$, the LCA is exact, and so \eqn{square1} is also superior
to \eqn{square2} for an exact evaluation.
For $n\geq6$, where the LCA has subleading corrections, it is best to 
take the leading terms from \eqn{square1}, and the subleading terms from
\eqn{square2} (these are the terms proportional to $a$ in 
\eqn{tree6square})~\cite{BeGiKu90}.  For $n=7$, ref.~\cite{BeGiKu90}
employed the linear dependences in the Kleiss-Kuijf relation to
find an even more compact form for the subleading terms than using
\eqn{square2}.  (See ref.~\cite{CMMP} for another approach to 
numerically evaluating multi-parton amplitudes beyond the LCA.)


\section{One-Loop Color Decompositions}
\label{OneLoopSection}

The standard color decomposition for one-loop $n$-gluon amplitudes in
$SU(N_c)$ gauge theory with $\nf$ flavors of quarks is~\cite{BKLoopColor}
\begin{eqnarray}
  \A_{n}^\oneloop = g^n &\biggl[&
N_c \sum_{\si\in S_n/{\Bbb Z}_n}
   \tr (\lambda^{a_{\si_1}}\cdots \lambda^{a_{\si_n}})
    A_{n;1}^{[1]}(\si_1,\ldots,\si_n)  \nonumber\\
&+& \hskip-4mm
  \sum_{c=2}^{\lfloor n/2\rfloor+1}
\sum_{\si \in S_n/S_{n;c}} 
 \tr (\lambda^{a_{\si_1}}\cdots \lambda^{a_{\si_{c-1}}})
 \tr (\lambda^{a_{\si_c}} \cdots  \lambda^{a_{\si_n}})
A_{n;c} (\si_1,\ldots,\si_n)
\nonumber\\
&+& \nf \sum_{\si \in S_n/{\Bbb Z}_n}
 \tr (\lambda^{a_{\si_1}}\cdots \lambda^{a_{\si_n}})
    A_{n;1}^{[1/2]}(\si_1,\ldots,\si_n) \biggr] \, ,
\label{LoopColor}
\end{eqnarray}
where $A_{n;c}$ are the subamplitudes,
$S_{n;c}$ is the subset of $S_n$ that leaves the double
trace structure invariant, and $\lfloor x \rfloor$ is the greatest 
integer less than or equal to $x$.  The superscript $[j]$ denotes
the spin of the particle circulating in the loop.  The subamplitudes
$A_{n;1}^{[j]}$ are color-ordered, and give rise to the leading
contributions to the cross-section in the large-$N_c$ limit.

The subleading subamplitudes $A_{n;c>1}$ can be obtained from the leading
ones $A^{[1]}_{n;1}$ through the permutation sum~\cite{BDDKNeq4}
\begin{equation}
A_{n;c>1}(1,2,\ldots,c-1;c,c+1,\ldots,n)\ =\
 (-1)^{c-1} \sum_{\si\in \COP\{\alpha\}\{\beta\}}
 A^{[1]}_{n;1}(\si_1,\ldots,\si_n) \, ,
\label{sublanswer}
\end{equation}
where $\alpha_i \in \{\alpha\} \equiv \{c-1,c-2,\ldots,2,1\}$,
$\beta_i \in \{\beta\} \equiv \{c,c+1,\ldots,n-1,n\}$,
and $\COP\{\alpha\}\{\beta\}$ is the set of all
permutations of $\{1,2,\ldots,n\}$ with $n$ held fixed
that preserve the cyclic
ordering of the $\alpha_i$ within $\{\alpha\}$ and of the $\beta_i$
within $\{\beta\}$, while allowing for all possible relative orderings
of the $\alpha_i$ with respect to the $\beta_i$.

\begin{figure}[ht]
\begin{center}
\begin{picture}(350,120)(0,0)
\Text(0,60)[r]{{\Large (a)}}
\CArc(80,60)(40,0,360)
\Line(25,60)(40,60) \Line(120,60)(135,60)
\Line(80,100)(80,115) \Line(80,20)(80,5)
\Line(114.641,80)(127.631,87.5)  \Line(100,94.641)(107.5,107.631)
\Line(60,94.641)(52.5,107.631)  \Line(45.359,80)(32.369,87.5)
\Line(45.359,40)(32.369,32.5)   \Line(60,25.359)(52.5,12.369)
\Line(100,25.359)(107.5,12.369)  \Line(114.641,40)(127.631,32.5)
\Text(200,60)[r]{{\Large (b)}}
\Text(224,60)[r]{{\Large $\bar{q}$}}
\Text(339,60)[l]{{\Large $q$}}
\CArc(280,60)(40,0,360)
\Line(280,100)(280,115) \Line(280,20)(280,5)
\Line(314.641,80)(327.631,87.5)  \Line(300,94.641)(307.5,107.631)
\Line(260,94.641)(252.5,107.631)  \Line(245.359,80)(232.369,87.5)
\Line(245.359,40)(232.369,32.5)   \Line(260,25.359)(252.5,12.369)
\Line(300,25.359)(307.5,12.369)  \Line(314.641,40)(327.631,32.5)
\SetWidth{1.5}
\ArrowArc(280,60)(40,180,360)
\Line(225,60)(240,60) \Line(320,60)(335,60)
\end{picture}
\end{center}
\caption[a]{\small (a) Ring color factors for one-loop $n$-gluon
amplitudes. (b) The corresponding color factors for one-loop amplitudes
with an external quark-antiquark pair.}
\label{RingFigure}
\end{figure}
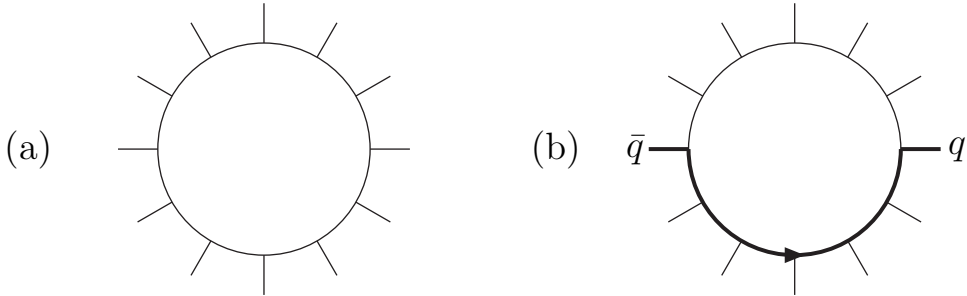

In a dual representation of one-loop amplitudes, the subleading 
subamplitudes $A_{n;c>1}(\{\alpha^T\};\{\beta\})$ come from the annulus
diagram where gluons belonging to $\{\alpha\}$ and $\{\beta\}$ are
emitted from the inner and outer boundaries of the annulus, respectively.
Thus their color properties are very similar to those of the tree 
subamplitudes $A_n^\tree(1,\{\alpha\},n,\{\beta\})$ if legs $1$ and $n$ 
are `sewn together', i.e. contracted with $\delta^{a_1a_n}$.
Indeed, with this interpretation, \eqn{sublanswer} has a very similar 
structure to the Kleiss-Kuijf relation~(\ref{KKequation}).
These remarks suggest that the new tree-level color decomposition
should have a one-loop analog, where the strings of structure constants
that appear are `ring diagrams' instead of multi-peripheral diagrams,
as depicted in \fig{RingFigure}a.

The new one-loop $n$-gluon color decomposition, expressed in terms 
of adjoint generator matrices $(F^a)_{bc}$, is
\begin{eqnarray}
\A_{n}^\oneloop = g^n 
 \sum_{\si\in S_{n-1}/{\cal R}}
   \Bigl[ \tr (F^{a_{\si_1}} \cdots F^{a_{\si_n}})
A_{n;1}^{[1]}(\si_1,\ldots,\si_n)
 + 2 \nf  \tr (\lambda^{a_{\si_1}} \cdots \lambda^{a_{\si_n}})
    A_{n;1}^{[1/2]}(\si_1,\ldots,\si_n) \Bigr].
\label{NewLoopColor}
\end{eqnarray}
Here $S_{n-1} \equiv S_n/{\Bbb Z}_n$ is the group of non-cyclic
permutations, and ${\cal R}$ is the reflection:  
${\cal R}(1,2,\ldots,n) = (n,\ldots,2,1)$.
Thus the number of linearly independent subamplitudes is $(n-1)!/2$.

The terms proportional to $\nf$ in \eqn{NewLoopColor} follow from
\eqn{LoopColor} using only the reflection identity satisfied by the
$A_{n;1}$:  $A_{n;1}^{[j]}(n,\ldots,2,1) 
= (-1)^n  \, A_{n;1}^{[j]}(1,2,\ldots,n)$.
To prove that the remaining terms in \eqn{NewLoopColor} are
correct, we follow closely the second proof of \eqn{GluonDecompNew}.
Using the Jacobi identity, the generic one-loop $f^{abc}$ color factor can
be turned into ring diagrams.  The manipulations are exactly the same as
those shown in \fig{TreeSimpFigure}, once legs $1$ and $n$ are joined
together to form a loop.  This establishes \eqn{NewLoopColor}, but with an
unknown coefficient for $\tr(F^{a_{\si_1}} \cdots F^{a_{\si_n}})$.  Next
we equate this expression to the standard color
decomposition~(\ref{LoopColor}), and contract both sides with
$\tr(\lambda^{a_{\si_1}} \cdots \lambda^{a_{\si_n}})$.  Again a unique
term survives on each side of the equation, at leading order in $N_c$, and
this establishes that the coefficient is correctly given by
$A_{n;1}^{[1]}(\si_1,\ldots,\si_n)$.  (If we had not removed the 
reflections ${\cal R}$ from \eqn{NewLoopColor}, then both a ring
diagram and its reflection would have contributed to the contraction;
this would have led to a factor of $\hf$ in the coefficient of 
$\tr(F^{a_{\si_1}} \cdots F^{a_{\si_n}})$.)
 
\Eqn{NewLoopColor} puts the gluon and fermion loops manifestly on the same
footing.  The contributions of the subleading-color subamplitudes
$A_{n;c>1}$ are neatly packaged into the ring-diagram color structures.
In addition, all of the generalized $U(1)$ decoupling identities
are automatically incorporated.  Finally, \eqn{NewLoopColor} shows
explicitly the difference between the tree-level and the one-loop color
decomposition of the gluon sector for any number of external legs. For
instance, if we take the amplitude for a Higgs boson plus three gluons,
both \eqn{TreeGluonColorOld} and the gluon-loop contribution to 
\eqn{LoopColor} have as a color factor 
$\tr(\lambda^{a_{\si_1}} \lambda^{a_{\si_2}} \lambda^{a_{\si_3}})$;
the actual difference in the color structure is concealed in the different
properties of the subamplitudes $A_3^\tree$ and $A_{3;1}^{[1]}$. 
In contrast, \eqn{GluonDecompNew} and \eqn{NewLoopColor} show explicitly 
the difference in color structure.

One-loop amplitudes contribute to NLO QCD
cross-sections through their interference with tree amplitudes.
Carrying out the color-summed NLO interference terms for amplitudes 
color-decomposed as in \eqns{GluonDecompNew}{NewLoopColor} is 
straightforward:
\begin{eqnarray}
\lefteqn{
\sum_{\rm colors}  \A(1,\dots,n) [\A(1,\dots,n)]^* 
{\Large\vert}_{\rm NLO} } \nonumber\\ && = 2\, \sum_{\rm colors} 
\Re \left( \A^\tree  (\A^\oneloop)^* \right) 
\label{nlosquare}\\ && = 2 (g^2)^{n-1}\, \Re \sum_{i=1}^{(n-2)!} 
\sum_{j=1}^{(n-1)!/2} A_i^\tree \left[ \hat{c}_{ij}  (A_j^{[1]})^* 
+ 2 \, \nf \, \hat{d}_{ij}  (A_j^{[1/2]})^* \right] \,
,\nonumber
\end{eqnarray}
where
\begin{eqnarray}
\hat{c}_{ij} &=& \sum_{\rm colors} \left(P_i \{F^{a_2} 
\cdots F^{a_{n-1}}\} \right)_{a_1 a_n}
\left[\tr \left(F^{a_1} P_j \{F^{a_2} \cdots F^{a_n}\}\right)\right]^*
\,, \nonumber\\
\hat{d}_{ij} &=& \sum_{\rm colors} \left(P_i \{F^{a_2} 
\cdots F^{a_{n-1}}\} \right)_{a_1 a_n} 
\left[\tr (\lambda^{a_1} P_j \{ \lambda^{a_2} \cdots \lambda^{a_n} \} 
)\right]^* ,\label{chat}
\end{eqnarray}
with $P_i$ the $i^{\rm th}$ permutation in $S_{n-2}$ and 
$P_j$ the $j^{\rm th}$ permutation in $S_{n-1}/{\cal R}$.

NNLO production rates include the 
virtual contributions,
\begin{equation}
\sum_{\rm colors}  \A(1,\dots,n) [\A(1,\dots,n)]^* 
{\Large\vert }_{\rm NNLO} = 
\sum_{\rm colors} \left[ \, 2\, \Re \left( \A^\tree  
(\A^\twoloop)^* \right) + |\A^\oneloop|^2 \,\right]\, 
.\label{nnlosquare}
\end{equation}
For the square of one-loop $n$-gluon amplitudes on the right-hand side 
of \eqn{nnlosquare}, we obtain
\begin{eqnarray}
\sum_{\rm colors} |\A^\oneloop|^2 &=& (g^2)^n\,\sum_{i,j=1}^{(n-1)!/2} 
\biggl[ \bar{c}_{ij} A_i^{[1]} (A_j^{[1]})^* 
+ 4 \, \nf \, \Re \,
  \left[ \bar{d}_{ij} A_i^{[1]}(A_j^{[1/2]})^* \right]  \nonumber\\ 
&& \hskip 2.5 cm 
+ \, 4 \, \nf^2 \, c_{ij} A_i^{[1/2]} (A_j^{[1/2]})^*  \biggr]\,,
\label{1loopsq}
\end{eqnarray}
where
\begin{eqnarray}
\bar{c}_{ij} &=&  
\sum_{\rm colors} \tr \left(F^{a_1} P_i \{F^{a_2} \cdots 
F^{a_n}\}\right) \left[\tr \left(F^{a_1} P_j \{F^{a_2} \cdots F^{a_n}\}
\right)\right]^* , \nonumber\\
\bar{d}_{ij} &=& 
\sum_{\rm colors} \tr \left(F^{a_1} P_i \{F^{a_2} \cdots F^{a_n}\}\right) 
\left[\tr (\lambda^{a_1} P_j \{ \lambda^{a_2} \cdots \lambda^{a_n} \} 
)\right]^* , 
\label{cbar} 
\end{eqnarray}
$c_{ij}$ is given in \eqn{cdef}, and $P_i$ is the $i^{\rm th}$ permutation 
in $S_{n-1}/{\cal R}$.  In  \app{sec:app1} we give the explicit values 
for the color matrices $\hat{c}_{ij}$, $\hat{d}_{ij}$, $\bar{c}_{ij}$, 
$\bar{d}_{ij}$ and $c_{ij}$ that are required for cross-section 
computations up to $\Ord(\alpha_s^4)$.

A similar analysis can be applied to one-loop amplitudes with an external
quark-antiquark pair plus $n-2$ gluons.  These amplitudes have the
standard color decomposition~\cite{TwoqNgluon}, 
\begin{eqnarray}
  \A_{n}^\oneloop(\qb_1,q_2,g_3,\ldots,g_n) = 
g^n  \sum_{j=1}^{n-1} \sum_{\si\in S_{n-2}/S_{n;j}}
   \Gr^{(\qb q)}_{n;j} (\si_3,\ldots,\si_n) 
               A_{n;j} (1_{\qb},2_q;\si_3,\ldots,\si_n),
\label{LoopColorqq}
\end{eqnarray}
where the color structures $\Gr^{(\qb q)}_{n;j}$ are defined by
\begin{eqnarray}
\Gr^{(\qb q)}_{n;1}(3,\ldots,n) &=& 
N_c (\lambda^{a_3} \cdots \lambda^{a_n})^{~\bar{i}_1}_{i_2} \,, \nonumber\\
\Gr^{(\qb q)}_{n;2}(3;4,\ldots,n) &=& 0, \nonumber\\
\Gr^{(\qb q)}_{n;j}(3,\ldots,j+1;j+2,\ldots,n) &=&
 \tr(\lambda^{a_3} \cdots \lambda^{a_{j+1}})
  (\lambda^{a_{j+2}} \cdots \lambda^{a_n})^{~\bar{i}_1}_{i_2} \,, 
  \qquad j=3,\ldots,n-2, \nonumber\\
\Gr^{(\qb q)}_{n;n-1}(3,\ldots,n) &=& 
  \tr(\lambda^{a_3} \cdots \lambda^{a_n}) \, \delta^{~\bar{i}_1}_{i_2} \,,
\label{Grqqdef}
\end{eqnarray}
and $S_{n;j} = {\Bbb Z}_{j-1}$ is the subgroup of $S_{n-2}$ that leaves
$\Gr^{(\qb q)}_{n;j}$ invariant.  

The leading-color subamplitudes in \eqn{LoopColorqq} can be expressed
in terms of color-ordered (`primitive') amplitudes as
\begin{eqnarray}
  \A_{n;1}(1_\qb,2_q,3,\ldots,n) &=& 
    A_n^{L,[1]}(1_\qb,2_q,3,\ldots,n) 
  - {1\over N_c^2} A_n^{R,[1]}(1_\qb,2_q,3,\ldots,n) \nonumber\\
 && \quad + {\nf \over N_c} A_n^{L,[1/2]}(1_\qb,2_q,3,\ldots,n),
\label{qqleadanswer}
\end{eqnarray}
whereas the subleading contributions are given by the permutation
formula~\cite{TwoqNgluon},
\begin{eqnarray}
\lefteqn{ 
 A_{n;j}(1_\qb,2_q;3,\ldots,j+1;j+2,j+3,\ldots,n) } \nonumber\\
  && = (-1)^{j-1} \sum_{\si\in\COP\{\alpha\}\{\beta\}} \biggl[
    A_n^{L,[1]}(\si(1_\qb,2_q,3,\ldots,n))
  - {\nf \over N_c} A_n^{R,[1/2]}(\si(1_\qb,2_q,3,\ldots,n)) \biggr] \,,
\label{qqsublanswer}
\end{eqnarray}
where $\{\alpha\} = \{ j+1,j,\ldots,4,3 \}$,
$\{\beta\} = \{ 1,2,j+2,j+3,\ldots,n-1,n \}$, with $1_\qb$ held fixed.  
In \eqns{qqleadanswer}{qqsublanswer}, the quantities $A_n^{L,[j]}$ and
$A_n^{R,[j]}$ are defined in terms of color-ordered Feynman diagrams,
much like the $n$-gluon subamplitudes $A_{n;1}^{[j]}$.  However, because
an external $\qb q$ pair is present now, one also has to specify which 
way the line running from the antiquark to the quark turns when it enters
the loop, left or right; this accounts for the additional index, $L$ or
$R$.  Again the index $[j]$ denotes the spin $j$ of a particle circulating
around the loop, for those graphs in the primitive amplitude where the
external $\qb q$ line does not enter the loop.  The reflection identity,
\begin{equation}
A_n^{R,[j]}(1_\qb,3,4,\ldots,2_q,\ldots,n-1,n) =
(-1)^n \, A_n^{L,[j]}(1_\qb,n,n-1,\ldots,2_q,\ldots,4,3),
\label{qqreflection}
\end{equation}
relates the left and right subamplitudes.

There are two types of subleading-color contributions in 
\eqn{qqsublanswer}, distinguished by whether they contain a factor of 
$\nf$ or not.  Although both types are generated by cyclicly ordered
permutations, their origins are quite different.  In particular, the
subleading $\nf$ terms arise from the $1/N_c$ term in the $SU(N_c)$
Fierz identity,
\begin{equation}
\sum_a (X_1 \lambda^a X_2) (Y_1 \lambda^a Y_2)
 = (X_1 Y_2) (Y_1 X_2) - {1\over N_c} (X_1 X_2) (Y_1 Y_2),
\label{sunFierz}
\end{equation}
where $X_i$, $Y_i$ are strings of generator matrices $\lambda^{a_i}$.
Such manifest $1/N_c$ terms are not generated by strings of structure
constants.  (In fact, $1/N_c$ terms are generally suppressed when
$f^{abc}$s are present, because $1/N_c$ factors come from explicitly
projecting out the $U(1)$ factor in $U(N_c) = SU(N_c) \times U(1)$, while
structure constants accomplish this projection automatically.)  Because of
this fact, the freedom to include structure constants in a color
decomposition does not seem to lead to any simpler representation of the
subleading $\nf$ terms in \eqn{qqsublanswer}, and we shall be content to
simplify the no-$\nf$ terms\footnote{
Since the $\nf$ terms in the $A_{n;j>1}$ subamplitudes arise from 
projecting out a $U(1)$ factor, they can also be viewed as subamplitudes 
for a fictitious $SU(N_1) \times SU(N_2) \times U(1)$ theory containing 
two fermion representations, say $({\bf N_1},{\bf 1})_{+1}$ (the external
$\qb q$ pair) and $({\bf 1},{\bf N_2})_{+1}$ (the fermion in the loop).
This description makes it a bit more transparent how the color structure
$\tr(\lambda^{a_3} \cdots \lambda^{a_{j+1}})$
$\times \, (\lambda^{a_{j+2}} \cdots \lambda^{a_n})^{~\bar{i}_1}_{i_2}$
multiplying these subamplitudes corresponds to their definition.  
But it does not lead to a distinct color decomposition.}.

The new color decomposition for one-loop amplitudes with an external
$\qb q$ pair is
\begin{eqnarray}
\A_{n}^\oneloop(\qb_1,q_2,g_3,\ldots,g_n)
&=& g^n \biggl[
\sum_{p=2}^n \sum_{\si \in S_{n-2}} 
(\lambda^{x_2} \lambda^{a_{\si_3}} \cdots \lambda^{a_{\si_p}}
 \lambda^{x_1})^{~\bar{i}_1}_{i_2}
(F^{a_{\si_{p+1}}} \cdots F^{a_{\si_n}})_{x_1 x_2} \nonumber\\
&& \hskip 2 cm
\times A_n^{R,[1]}(1_\qb,\si_{p+1},\ldots,\si_n,2_q,\si_3,\ldots,\si_p)
\label{LoopColorqqNew}\\
&+& {n_f\over N_c} \, \sum_{j=1}^{n-1} \sum_{\si\in S_{n-2}/S_{n;j}}
   \Gr^{(\qb q)}_{n;j} (\si_3,\ldots,\si_n) 
               A_{n;j}^{[1/2]} (1_{\qb},2_q;\si_3,\ldots,\si_n) \biggr]\,
,\nonumber
\end{eqnarray}
where for $p=n$ the product of generators in the adjoint representation
reduces to the identity, $(F \cdots F)_{x_1x_2} \to \delta_{x_1x_2}$;
the color structures $\Gr^{(\qb q)}_{n;j}$ are defined in \eqn{Grqqdef};
and
\begin{eqnarray}
A_{n;1}^{[1/2]}(1_\qb,2_q,3,\ldots,n) 
&=& A_n^{L,[1/2]}(1_\qb,2_q,3,\ldots,n),
\label{newqqsubamp} \\
A_{n;j}^{[1/2]}(1_\qb,2_q;3,\ldots,j+1;j+2,j+3,\ldots,n) 
  &=& (-1)^j \sum_{\si\in\COP\{\alpha\}\{\beta\}} 
A_n^{R,[1/2]}(\si(1_\qb,2_q,3,\ldots,n))\, ,\nonumber
\end{eqnarray}
with $\{\alpha\}$ and $\{\beta\}$ as in \eqn{qqsublanswer}.
The new types of color structures appearing in \eqn{LoopColorqqNew} are 
shown in \fig{RingFigure}b.  

Most of the subamplitudes $A_n^{R,[1]}$ do not contribute at leading order
in the large $N_c$ limit.  (Only those of the form
$A_n^{R,[1]}(1_\qb,\ldots,2_q)$ do.)  Therefore the technique of
contracting with a suitable color structure at large $N_c$ cannot be used
to prove \eqn{LoopColorqqNew}.  Instead we follow the first line of
reasoning that we used at tree-level; that is, we substitute for the
structure constants in terms of $\lambda$ matrices and collect terms.  The
result agrees with \eqn{LoopColorqq}, after substituting in
\eqns{qqleadanswer}{qqsublanswer} for the leading and subleading
subamplitudes and making use of the reflection
identity~(\ref{qqreflection}).

The contribution of one-loop amplitudes with an external quark-antiquark
pair to NLO QCD cross sections parallels the $n$-gluon case, \eqn{nlosquare}.
Carrying out the color-summed NLO interference terms for amplitudes 
color-decomposed as in \eqns{QuarkDecompOld}{LoopColorqqNew}, and 
omitting the $\nf$ terms, we obtain,
\begin{eqnarray}
\lefteqn{
\sum_{\rm colors}  \A(\qb_1,q_2,g_3,\ldots,g_n)_{{\rm no}-\nf} 
[\A(\qb_1,q_2,g_3,\ldots,g_n)_{{\rm no}-\nf}]^* 
{\Large\vert}_{\rm NLO} } \nonumber\\ && = 2\, \sum_{\rm colors} 
\Re \left( \A^\tree  (\A^\oneloop)^*_{{\rm no}-\nf} \right) 
\label{nloqqsquare}\\ && = 2 (g^2)^{n-1}\, \Re \sum_{p=2}^n 
\sum_{i,j=1}^{(n-2)!} \hat{m}_{ij}^{(p)} 
A_i^\tree(1_\qb,2_q,i_3,\ldots,i_n)   
\left[A_j^{R,[1]}(1_\qb,j_{p+1},\ldots,j_n,2_q,j_3,\ldots,j_p)\right]^* \,
,\nonumber
\end{eqnarray}
where
\begin{equation}
\hat{m}_{ij}^{(p)} = \sum_{\rm colors} \left( P_i
\{\lambda^{a_3} \cdots \lambda^{a_n}\}\right)^{~\bar{i}_1}_{i_2}
\left[ P_j \{ (\lambda^{x_2} \lambda^{a_3} \cdots \lambda^{a_p}
 \lambda^{x_1})^{~\bar{i}_1}_{i_2} (F^{a_{p+1}} \cdots F^{a_n})_{x_1 x_2}
\} \right]^\dagger\, ,\label{mhat}
\end{equation}
with $P_i$ the $i^{\rm th}$ permutation in $S_{n-2}$, and
$P_j$ the $j^{\rm th}$ permutation in $S_{n-2}$ at fixed $p$.
The partitions $p$ span a matrix-valued vector $\hat{m}_{ij}^{(p)}$,
which we give in appendix A for $n=4,5$,
as required for cross-section computations up to $\Ord(\alpha_s^4)$.
For $n=5$, we have checked that \eqn{nloqqsquare}, with $\hat{m}_{ij}^{(p)}$
given in \eqn{nloqqgggsum}, agrees with the no-$\nf$ terms of 
ref.~\cite{TwoqNgluon}.
The $\nf$ terms for $n=4,5$ have been computed in refs.~\cite{kst} and
\cite{TwoqNgluon}, respectively.

Finally, we note that by converting gluons into photons we can obtain
multi-photon amplitudes at tree and loop level.  In order to convert a
gluon into a photon we make use of the embedding of $U(1)$ in $U(N_c)$.
We replace the quark-gluon vertex factor $g\lambda^a$ with the
quark-photon vertex factor $\sqrt{2} \, e Q_q \, I$, where $Q_q$ is the
quark electric charge and the factor $\sqrt{2}$ is due to our choice of
normalization of the $\lambda$ matrices.  Since the identity matrix $I$
commutes with the $\lambda$ matrices, all possible attachments of the
photon to the quark line contribute to the same color structure.  Thus
subamplitudes involving a photon can be obtained by summing over
permutations of gluon subamplitudes in which the photon assumes all
possible places in the ordering.

Photons do not couple to gluons. In a $f^{abc}$-based decomposition this
is apparent:  When the index $a$ corresponds to the identity matrix $I$,
the $U(N_c)$ generator $F^a$ in the adjoint representation vanishes.  In
this paper we presented new color decompositions for gluon tree
amplitudes, \eqn{GluonDecompNew}, and for the gluon-loop sector of
one-loop amplitudes, \eqns{NewLoopColor}{LoopColorqqNew}; we still use the
$\lambda$-based decomposition for quark-gluon tree amplitudes,
\eqn{QuarkDecompOld}, and for the quark-loop sector of one-loop
amplitudes.  Thus it might seem that we can bring no new insight to the
calculation of multi-photon amplitudes. However, that is not the case for
one-loop amplitudes with an external $\qb q$ pair: In the no-$\nf$ terms
in \eqn{LoopColorqqNew} a photon can still attach to the external quark
line, i.e. to the string of $\lambda$ matrices; however, it manifestly
cannot attach to the string of $F$ matrices.  
We immediately see that subamplitudes of type 
$A_n^{R,[1]}(1_\qb,\si_3,\ldots,\si_n,2_q)$,
corresponding in \eqn{LoopColorqqNew} to partitions with $p=2$, do not
contribute, because there is no $\lambda^a$ matrix to replace by the
identity.  (The $\lambda^{x_i}$ matrices correspond to internal gluons and
so should not be replaced.)  In the $\lambda$-based
decomposition~\cite{dkm}, the vanishing contribution of these
subamplitudes is realized only after summing over all permutations.  Thus
the new color decomposition we obtain for $\qb qg\cdots g\gamma$ one-loop
amplitudes, by converting a gluon into a photon in \eqn{LoopColorqqNew}, 
and summing over all the possible attachments of the photon to the quark 
line, is more compact than the one obtained from the $\lambda$-based 
decomposition.  For $n=5$ we verified that explicitly by computing the 
${\bar q}qgg\gamma$ one-loop amplitude from \eqn{LoopColorqqNew}, and 
comparing it to (and checking its equivalence with) the same amplitude, 
but $\lambda$-decomposed~\cite{dkm}.


\section{Discussion}
\label{ConclusionsSection}

In ref.~\cite{dfm} a new color decomposition was presented for the
$n$-gluon tree amplitude; it was shown to be equivalent to the standard
one up to $n=7$ and was conjectured to be valid for arbitrary $n$.  In the
present paper we have proven this conjecture.  We have also presented, and
proven, analogous color decompositions for one-loop $n$-gluon amplitudes
and one-loop amplitudes with an external quark-antiquark pair plus $n-2$
gluons.

The new color decompositions have the advantage of being written in
terms of just the independent subamplitudes.  This makes
the decomposition of an amplitude more compact.  It is particularly
useful for computing the color-summed interference terms between tree 
amplitudes and one-loop amplitudes, and the square of one-loop amplitudes,
which are relevant respectively for NLO and NNLO calculations of jet 
production rates.  The color-summed interference terms and squares are 
given in terms of color matrices; we have provided in the appendix 
explicit values required for cross-section computations up to 
$\Ord(\alpha_s^4)$.

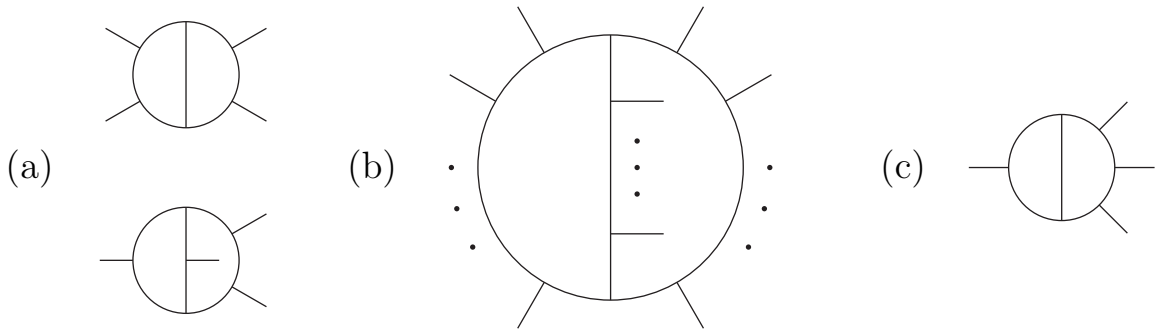
\begin{figure}[ht]
\begin{center}
\begin{picture}(420,170)(0,0)
\Text(0,85)[r]{{\Large (a)}}
\CArc(50,120)(20,0,360) \Line(50,100)(50,140)
\Line(67.3205,130)(80.3109,137.5)
\Line(67.3205,110)(80.3109,102.5)
\Line(32.6795,130)(19.6891,137.5)
\Line(32.6795,110)(19.6891,102.5)
\CArc(50,50)(20,0,360) \Line(50,30)(50,70)
\Line(67.3205,60)(80.3109,67.5)
\Line(67.3205,40)(80.3109,32.5)
\Line(30,50)(17.5,50) \Line(50,50)(62.5,50)
\Text(130,85)[r]{{\Large (b)}}
\CArc(210,85)(50,0,360) \Line(210,35)(210,135)
\Line(253.3013,110)(270.6218,120)
\Line(235,128.3013)(245,145.6218)
\Vertex(270,85){1} \Vertex(267.9555,69.4709){1} \Vertex(261.9615,55){1}
\Line(235,41.6987)(245,24.3782)
\Line(166.6987,110)(149.3782,120)
\Line(185,128.3013)(175,145.6218)
\Vertex(150,85){1} \Vertex(152.0444,69.4709){1} \Vertex(158.0385,55){1}
\Line(185,41.6987)(175,24.3782)
\Line(210,110)(230,110)
\Vertex(220,95){1} \Vertex(220,85){1} \Vertex(220,75){1}
\Line(210,60)(230,60) 
\Text(330,85)[r]{{\Large (c)}}
\CArc(380,85)(20,0,360) \Line(380,65)(380,105)
\Line(394.1421,99.1421)(404.7487,109.7487)
\Line(394.1421,70.8579)(404.7487,60.2513)
\Line(400,85)(415,85) \Line(360,85)(345,85)
\end{picture}
\end{center}
\caption[a]{\small (a) The only color factors appearing in the two-loop 
4-gluon amplitude in $N=4$ super-Yang-Mills theory. 
(b) The types of color factors for multi-loop $n$-gluon 
amplitudes that remain after removing all nontrivial trees.
(c) Example of an `unwanted' 4-gluon planar color factor.}
\label{MultiLoopFigure}
\end{figure}

We expect that structure-constant-based decompositions will also be
useful for general multi-loop scattering amplitude calculations, once they
have been derived.  In a particular case they have already proven useful:
Two-loop four-gluon amplitudes in $N=4$ supersymmetric Yang-Mills theory
were computed via their unitarity cuts and were presented in a trace-based
formalism~\cite{BRY}, but they can be more compactly
expressed in an $f^{abc}$ basis~\cite{BDDPR}.  In the latter basis only two
types of color structures appear; these are represented graphically in
\fig{MultiLoopFigure}a by the planar and nonplanar double box diagrams.
The kinematic object multiplying each structure is simply the scalar
integral (i.e., the $\phi^3$ Feynman diagram) for the graph with the same
topology.  

Some of this simplicity is undoubtedly due to $N=4$ supersymmetry.
However, group theory alone restricts the possible
multi-loop color structures considerably.  At one loop, we used the Jacobi
identity to remove nontrivial trees from the set of $n$-gluon color
structures.  We can similarly remove all such trees from multi-loop
$n$-gluon color structures.  The remaining color factors are illustrated
in \fig{MultiLoopFigure}b (for the two-loop case).  In the one-loop case,
this reduction exhausted the use of the Jacobi identity; in the multi-loop
case this is no longer true.  For example, in the two-loop four-gluon
amplitude the nonplanar double box diagram in \fig{MultiLoopFigure}a could
be removed in favor of the planar double box diagram in
\fig{MultiLoopFigure}a and another planar diagram, shown in
\fig{MultiLoopFigure}c.  Yet this would be an unwise move, at least in the
$N=4$ supersymmetric case, in the sense that the nonplanar scalar double
box integral would then have to be associated with planar color factors.
Thus the Jacobi identity should be wielded more judiciously in the
multi-loop case, and we leave this to future work.

Finally, we remark that even if a general, multi-loop $f^{abc}$-based
color decomposition is not yet available, the results in this paper can
still be employed in the context of specific multi-loop gluonic
calculations based on unitarity cuts~\cite{BRY,BDKReview,BDDPR}.  For
example, the unitarity cuts of two-loop amplitudes are given in terms of
tree-level and one-loop amplitudes.  Color decomposing these amplitudes as
in this paper leads directly, after sewing the color factors on each side
of the cut, to an $f^{abc}$-based decomposition for the cut of the
two-loop amplitude.  This decomposition is gauge invariant on the cut, by
virtue of the gauge invariance of the tree and one-loop amplitudes.  The
only potential difficulty is that to reconstruct the amplitude from its
cuts, one has to find a single function that consistently matches all the
cuts, and the color factors may have to be manipulated at this point.
Still, this seems likely to be a more efficient than a corresponding
approach using a trace-based decomposition of tree and one-loop
amplitudes, at least for computing terms that are subleading in the number
of colors.

At first sight it may seem like a step backward to abandon trace-based
color decompositions of gluonic scattering amplitudes in favor of ones
based on structure constants.  However, we have seen in this paper that 
with an appropriate choice of color factors, one does not really have to 
give up any desirable properties of the trace-based formalism (such as 
gauge invariance), at least at the tree and one-loop level.  We look 
forward to multi-loop extensions of this work, and their application to
more precise jet rate computations.


\vskip .3 cm 
\noindent
{\bf Acknowledgments}

We thank Zvi Bern and Michelangelo Mangano for useful conversations and
comments on the manuscript.


\appendix
\section{The color matrices}
\label{sec:app1}
We first give the color matrices appearing in the color-summed tree-level 
cross-section, defined in \eqn{ctilde}.  For $n=4$ they are
\begin{eqnarray}
\tilde{c}_{ij}={\cal C}_4(N_c) \cdot
\left(
\begin{array}{cc}
4 & 2 \\
2 & 4 
\end{array}
\right)
\;,\qquad
A^\tree_i=
\left(
\begin{array}{c}
(1,2,3,4) \\
(1,3,2,4) 
\end{array}
\right) \ ,
\end{eqnarray}
and for $n=5$, 
\begin{eqnarray}
\tilde{c}_{ij}={\cal C}_5(N_c) \cdot
\left(
\begin{array}{cccccc}
 8& 4& 4& 2& 2& 0\\ 
 4& 8& 2& 0& 4& 2\\ 
 4& 2& 8& 4& 0& 2\\ 
 2& 0& 4& 8& 2& 4\\ 
 2& 4& 0& 2& 8& 4\\ 
 0& 2& 2& 4& 4& 8
\end{array}
\right)
\;,\qquad
A^\tree_i=
\left(
\begin{array}{c}
(1, 2, 3, 4, 5)\\  
(1, 2, 4, 3, 5)\\  
(1, 3, 2, 4, 5)\\  
(1, 3, 4, 2, 5)\\  
(1, 4, 2, 3, 5)\\ 
(1, 4, 3, 2, 5)
\end{array}
\right)\ ,
\end{eqnarray}
where ${\cal C}_n(N_c)$ is defined in \eqn{calc}.
The above results agree with ref.~\cite{BeGiKu90},
up to an overall factor due to the different choice of normalization
of the matrices in the fundamental representation. 
The color matrix $\tilde{c}_{ij}$ for $n>5$  shows the $\Ord(1/N_c^2)$
dependence outlined in \eqn{square2}.  
For $n=6$, $\tilde{c}_{ij}$ is a $24\times 24$ matrix.  Rather than
give the entire matrix, we note that its essential information is contained
in its first row, i.e. in the matrix elements of $(1,2,3,4,5,6)$ with 
the 24 permutations.  Defining
$\tilde{c}_{(123456)(1\si(2345)6)} 
\equiv {\cal C}_6(N_c) \, \gamma(\si(2345))$,
we have
\begin{eqnarray}
\begin{array}{cccc}
\g(2345) = 16,& \qq \g(3245) = 8, & \qq \g(4235) = 4,& \qq \g(5234) = 2,\\
\g(2354) =  8,& \qq \g(3254) = 4, & \qq \g(4253) = 2,& \qq \g(5243) = 0,\\
\g(2435) =  8,& \qq \g(3425) = 4, & \qq \g(4325) = 0,& \qq \g(5324) = 0,\\
\g(2453) =  4,& \qq \g(3452) = 2, & \qq \g(4352) = 0,& \qq \g(5342) = a,\\
\g(2534) =  4,& \qq \g(3524) = 2, & \qq\g(4523) = 2+a,&\qq \g(5423) = a,\\
\g(2543) =  0,& \qq \g(3542) = 0, & \qq \g(4532) = a,& \qq \g(5432) = a,\\
\end{array}
\label{tree6square}
\end{eqnarray}
where $a=24/N_c^2$.  The terms proportional to $a$ agree with eq.~(A.10)
of ref.~\cite{BeGiKu90}.

The color matrices for the tree-level one-loop interference, defined in
\eqn{chat}, are for $n=4$
\begin{eqnarray}
&&\hat{c}_{ij}={\cal C}_5(N_c) \cdot
\left(
\begin{array}{ccc}
2 & -2 & 0\\ 
0 & -2 & 2
\end{array}
\right)
\;,\qquad
\hat{d}_{ij}={\cal C}_4(N_c) \cdot
\left(
\begin{array}{ccc}
1 & -1 & 0\\ 
0 & -1 & 1
\end{array}
\right) \;, \\
&&(A^{[1]}_{4;1})_j=
\left(
\begin{array}{c}
 (1, 2, 3, 4)\\
 (1, 2, 4, 3)\\
 (1, 3, 2, 4)
\end{array}
\right) \;,
\label{chat5}
\end{eqnarray}
and for $n=5$,
\begin{eqnarray}
&&\hat{c}_{ij}={\cal C}_6(N_c) \cdot
\left(
\begin{array}{cccccccccccc}
 2& -2& 0& -2& 0& 2& 0& 0& a& -a& a& a\\ 
 0& -2& 2& -2& 2& 0& a& a& a& 0& 0& a\\ 
 0& 0& a& -2& -a& 2& 2& -2& 0& 0& a& a\\ 
 a& a& a& 0& 0& 2& 0& -2& 2& 2& a& 0\\ 
 a& -2& 0& 0& 2& -a& a& 0& a& 2& 2& 0\\ 
 a& 0& a& a& 2& 0& a & -2& 0& 2& 0& 2
\end{array}
\right) \;, \\[10pt]
&&\hat{d}_{ij}={\cal C}_5(N_c) \cdot
\left(
\begin{array}{cccccccccccc}
 1& -1& 0& -1& 0& 1& 0& 0& b& -b& b& b\\ 
 0& -1& 1& -1& 1& 0& b& b& b& 0& 0& b\\ 
 0& 0& b& -1& -b& 1& 1& -1& 0& 0& b& b\\ 
 b& b& b& 0& 0& 1& 0& -1& 1& 1& b& 0\\ 
 b& -1& 0& 0& 1& -b& b& 0& b& 1& 1& 0\\ 
 b& 0& b& b& 1& 0& b& -1& 0& 1& 0& 1
\end{array}
\right) \;, \\
&&(A^{[1]}_{5;1})_j=
\left(
\begin{array}{c}
(1, 2, 3, 4, 5)\\ (1, 2, 3, 5, 4)\\ (1, 2, 4, 3, 5)\\ 
   (1, 2, 4, 5, 3)\\ (1, 2, 5, 3, 4)\\ (1, 2, 5, 4, 3)\\ 
   (1, 3, 2, 4, 5)\\ (1, 3, 2, 5, 4)\\ (1, 3, 4, 2, 5)\\ 
   (1, 3, 5, 2, 4)\\ (1, 4, 2, 3, 5)\\ (1, 4, 3, 2, 5)
\end{array}
\right) \;,
\label{a51def}
\end{eqnarray}
where $a=24/N_c^2$, $b=2/N_c^2$.
Although the matrices $\hat{c}_{ij}$ and $\hat{d}_{ij}$ have a relatively
small dimension, they are not particularly sparse.  We can improve the
sparseness considerably by using the tree-level $U(1)$ decoupling identity,
\eqn{DecouplingId}, in every column where three identical entries appear, 
thus trading three terms for one.
Of course, more than $(n-2)!$ different tree subamplitudes will now appear.
Using also the tree-level reflection identity, we arrive at the formula
\begin{eqnarray}
\lefteqn{
 \sum_{\rm colors}  \A(1,\dots,5) [\A(1,\dots,5)]^* 
{\Large\vert}_{\rm NLO} } \label{nlo5final} \\
&& = 4 g^8 {\cal C}_6(N_c) \, \Re \sum_{i=1}^{(n-1)!/2}
 \biggl[ A_i^\tree 
         \Bigl( A_i^{[1]} + {\nf \over N_c} \, A_i^{[1/2]} \Bigr)^*
 + {1\over N_c^2} A_{(24135)\cdot i}^\tree 
         \Bigl( 12 A_i^{[1]} + 2 \, {\nf \over N_c} \, A_i^{[1/2]} \Bigr)^* 
 \biggr] \,,
\nonumber
\end{eqnarray}
where $(24135)\cdot i$ is the permutation $(24135)$ composed with the
$i^{\rm th}$ permutation.  In comparing this formula to eq.~(11) of
ref.~\cite{FiveGluon}, we see that the $A_{5;3}$ terms can be removed, at
the price of multiplying the remaining $(1/N^2)$-suppressed pure-glue (no
$\nf$) terms by a factor of six.  This form for the NLO five-gluon
cross-section has been noted independently by
W. Kilgore~\cite{KilgoreUnpub}.

For the NNLO terms in \eqn{1loopsq} with $n=4$ we have  
\begin{eqnarray}
&&\bar{c}_{ij}={\cal C}_6(N_c) \cdot
\left(
\begin{array}{ccc}
2 + a& a& a\\
a& 2 + a& a\\
a& a& 2 + a\\
\end{array}
\right) \;,
\end{eqnarray}
\begin{eqnarray}
&&\bar{d}_{ij}={\cal C}_5(N_c) \cdot
\left(
\begin{array}{ccc}
1 + b& b& b\\
b& 1 + b& b\\
b& b& 1 + b\\
\end{array}
\right) \;,
\end{eqnarray}
\begin{eqnarray}
&&c_{ij}={\cal C}_4(N_c) \cdot
\left(
\begin{array}{ccc}
  1-b+d& d& d\\
  d& 1-b+d& d\\
  d& d& 1-b+d\\
\end{array}
\right) \;,
\qquad
d = - \frac{1}{N_c^2}+ \frac{3}{N_c^4} \ ,
\end{eqnarray}
where $a,b$ are defined below \eqn{a51def}.  Equivalently,
\begin{eqnarray}
\lefteqn{
\sum_{\rm colors} |\A^\oneloop(1,2,3,4)|^2 } \nonumber\\
&&= g^8 \, {\cal C}_6(N_c) \, \biggl[ \, \sum_{i=1}^3 
\Bigl( 2 \, \big\vert A_i^{[1]} \big\vert^2 
     + 4 {\nf\over N_c} \, \Re \, A_i^{[1]} (A_i^{[1/2]})^*
     + 4 {\nf^2\over N_c^2} (1-b) \, \big\vert A_i^{[1/2]} \big\vert^2 \Bigr)
\nonumber\\
&& \hskip 0.5 cm
+ \, a \Big\vert \sum_{i=1}^3 A_i^{[1]} \Big\vert^2
+ 4 {\nf\over N_c} b \, \Re \, \Bigl( \sum_{i=1}^3 A_i^{[1]} \Bigr)
                         \Bigl( \sum_{i=1}^3 A_i^{[1/2]} \Bigr)^*
+ 4 {\nf^2\over N_c^2} \, d \, 
    \Big\vert \sum_{i=1}^3 A_i^{[1/2]} \Big\vert^2 \, \biggr]\,.
\label{nnlo4final}
\end{eqnarray}

The color matrix-valued vector $\hat{m}^{(p)}_{ij}$ for the
tree-level one-loop interference for NLO processes with an
external quark-antiquark pair, defined in \eqn{mhat}, is for
$n=4$,
\begin{eqnarray}
\hat{m}^{(2,3,4)}_{ij}={\cal C}_4(N_c)\;  \cdot \left(
\begin{array}{c}
\left(
\begin{array}{cc}
 0& 1 \\
 1& 0
\end{array}
\right) \\
\displaystyle{\frac{1}{N_c^2}} \left(
\begin{array}{cc}
 1& 1 \\
 1& 1
\end{array}
\right) \\
\displaystyle{\frac{1}{N_c^2} } \left(
\begin{array}{cc}
 e& 1+e \\
 1+e& e
\end{array}
\right)
\end{array}
\right) \label{nloqqggsum}
\end{eqnarray}
\begin{eqnarray}
A^\tree_i = \left(
\begin{array}{cc}
(1, 2, 3, 4)& (1, 2, 4, 3)\\
\end{array}
\right)\ ,
\end{eqnarray}
where $e=1/N_c^2$. For $n=5$ we obtain,
\vfill\eject
\begin{eqnarray}
&& \hat{m}^{(2,3,4,5)}_{ij}=-{\cal C}_5(N_c) \label{nloqqgggsum}\\
&& \hspace{-0.8cm} \times \left(
\begin{array}{c}
\left(
\begin{array}{cccccc}
  b&b&b&0&0&1\\
  b&b&0&1&b&0\\
  b&0&b&b&1&0\\
  0&1&b&b&0&b\\
  0&b&1&0&b&b\\
  1&0&0&b&b&b
\end{array}
\right)
\\
\displaystyle{\frac{1}{N_c^2} } \left(
\begin{array}{cccccc}
   0&1&2&1&0&1\\
   1&0&0&1&2&1\\
   2&1&0&1&1&0\\
   0&1&1&0&1&2\\
   1&2&1&0&0&1\\
   1&0&1&2&1&0
\end{array}
\right)
\\
\displaystyle{\frac{1}{N_c^2} } \left(
\begin{array}{cccccc}
  e&e&1 + e&e&-1 + e&1 + e\\
    e&e&-1 + e&1 + e&1 + e&e\\
    1 + e&e&e&e&1 + e&-1 + e\\
    -1 + e&1 + e&e&e&e&1 + e\\
    e&1 + e&1 + e&-1 + e&e&e\\
    1 + e&-1 + e&e&1 + e&e&e
\end{array}
\right)
\\
\displaystyle{\frac{1}{N_c^2} }
\left(
\begin{array}{cccccc}
   f&e + f&e + f&-1 + 2e + f&-1 + 2e + f& 3e + f\\
   e + f&f&-1 + 2e + f&3e + f&e + f&    -1 + 2e + f\\
    e + f&-1 + 2e + f&f&e + f&    3e + f&-1 + 2e + f\\
    -1 + 2e + f&3e + f&e + f&f&-1 + 2e + f&    e + f\\
    -1 + 2e + f&e + f&3e + f&    -1 + 2e + f&f&e + f \\
    3e + f&-1 + 2e + f&-1 + 2e + f&e + f&e + f& f
\end{array}
\right)
\end{array}
\right) \nonumber
\end{eqnarray}
\begin{eqnarray}
A^\tree_i = \Bigl( (1, 2, 3, 4, 5)\ (1, 2, 3, 5, 4)\ (1, 2, 4, 3,
5)\ (1, 2, 4, 5, 3)\ (1, 2, 5, 3, 4)\ (1, 2, 5, 4, 3) \Bigr)\ ,
\end{eqnarray}
where $f=1/N_c^4$.  In both \eqns{nloqqggsum}{nloqqgggsum}, for
each value of $p$, the permutations $P_j$ of the gluons in the
loop amplitude are understood to appear in the same order as the
tree amplitude permutations $A^\tree_i$.

\end{document}